\documentclass[aps,pra,floats,floatfix,twocolumn,longbibliography,superscriptaddress]{revtex4-1}

\usepackage{color} 
\usepackage{latexsym}
\usepackage{amsmath} 
\usepackage{amssymb} 
\usepackage{bm}
\usepackage{wasysym}

\usepackage{graphicx}
\usepackage{epstopdf}
\graphicspath{{./}{./Figs/}}

\usepackage[colorlinks,linkcolor=blue,urlcolor=blue,bookmarks=true,pdftitle={}]{hyperref}

\newcommand{\beq}{\begin{eqnarray}}
\newcommand{\eeq}{\end{eqnarray}} 

\newcommand{\hide}[1]{}

\begin{document}

\title{Temporal bistability in the dissipative Dicke-Bose-Hubbard system}

\author{Tianyi Wu}
\affiliation{Physikalisches Institut, Rheinische Friedrich-Wilhelms-Universit\"at Bonn, Nu\ss allee 12, 53115, Bonn, Germany}
\author{Sayak Ray}
\affiliation{Physikalisches Institut, Rheinische Friedrich-Wilhelms-Universit\"at Bonn, Nu\ss allee 12, 53115, Bonn, Germany}
\author{Johann Kroha} 
\affiliation{Physikalisches Institut, Rheinische Friedrich-Wilhelms-Universit\"at Bonn, Nu\ss allee 12, 53115, Bonn, Germany}
\affiliation{\hbox{School of Physics and Astronomy, University of St. Andrews, North Haugh, St. Andrews, KY16 9SS, United Kingdom}}

\begin{abstract}

We consider a driven-dissipative system consisting of an atomic Bose-Einstein condensates loaded into a two-dimensional Hubbard lattice and coupled to a single mode of an optical cavity. Due to the interplay between strong, repulsive atomic interaction and the atom-cavity coupling, the system exhibits several phases of atoms and photons including the atomic superfluid (SF) and supersolid (SS). We investigate the dynamical behaviour of the system, where we include dissipation by means of Lindblad master equation formalism. Due to the discontinuous nature of the Dicke transition for strong atomic repulsion, we find extended co-existence region of different phases. We investigate the resulting switching dynamics, particularly between the coexisting SF and SS phases, which eventually becomes damped by the dissipation. 

\end{abstract}
\maketitle


\section{Introduction}

Open quantum systems are inherent in nature since no system can completely be isolated from the environment. They are beyond the description by Schrödinger unitary time evolution and can give rise to various non-equilibrium phenomena like decoherence, dissipation, relaxation dynamics towards equilibrium, non-Markovian dynamics etc. \cite{petru}. 
Ultracold atomic systems turn out to be a versatile platform to study such non-equilibrium phenomena where dissipation can arise from the atom-number loss, spontaneous decay of atomic excitations, or coupling the condensate to cavity modes in an optical resonator, and the system parameters can be engineered appropriately \cite{Zoller12}.
This has generated an impetus to study the interplay of interaction and dissipation in a generic quantum many-body system.
For example, the effects of non-equilibrium and dissipative dynamics have been studied in 
Bose-Josephson junctions \cite{Stringari01, Posazhennikova16, Tim18}, in a spin-boson model where decoherence leads to termination of quantum tunneling \cite{Leggett81}, in  trapped-ion systems \cite{Zoller11} and in ultracold Rydberg atom gases with spontaneous decay \cite{Weimer10, Arimondo11, Ray16, Hofstetter19}. 
Recently, even dynamical vibrational modes of optical lattices via coupling to a dynamical resonator have been realized \cite{Esslinger13}, and it has opened up a new research line of cavity-QED with quantum gases \cite{Piazza21}.

A prominent case is the Dicke model \cite{Dicke54}. It exhibits a quantum phase transition from a normal state to a superradiant state of the atoms \cite{Brandes03} which has been realized in experiment by coupling the atomic condensate to a single mode of an optical cavity \cite{Esslinger10}. The cavity photons can mediate an effective, long-range interaction between the atoms, which leads to the realization of a supersolid phase of matter \cite{Landig16, Hemmerich15, Hofstetter13, Thorwart15, Kollath20}. The effect of non-coherent photon fluctuations has been taken into account in \cite{Rosch21}. 
The Dicke system has often been regarded as an equilibrium model where the superradiant transition is induced by increasing the atom-cavity coupling strength. However, taking into account realistic, driven-dissipative effects even in a stationary situation, where the atoms are excited by an external laser and photons are lost due to cavity imperfections \cite{Kessler15}, has started only in the most recent works \cite{Keeling19, Lesanovsky14, Kollath20, Rosch21, Ray22}. 

In the present paper, we study a driven-dissipative, interacting, atomic Bose gas loaded in a two-dimensional (2D) optical lattice and coupled to a single optical-cavity mode in the presence of cavity-photon loss. We focus on the time-dependent dynamics after an initial non-equilibrium excitation. We cover the entire atomic interaction range from weak to strong repulsion $U$, which includes Bose-Einstein condensation or Mott-Hubbard localization of the atom system, respectively. The resulting interplay of these low-energy phases with the cavity-photon system gives rise to a rich phase diagram. It exhibits Mott insulator (MI) or homogeneous, superfluid (SF) states, respectively, coexisting with the normal-radiant (NR) phase of the cavity-photon system, while the superradiant (SR) photon phase induces self-organized spatial modulation of the atom system, i.e., a normal density wave (DW) or a supersolid (SS) phase, depending on pump as well as interaction strength.  

We treat the cavity-photon field semi-classically, whereas the atomic condensate phases and dynamics in the Hubbard lattice are described using the Gutzwiller cluster mean-field theory \cite{Luehmann13, Pohl22}.  
Interestingly, we find that the Dicke normal-to-superradiant transition, especially in the strong coupling limit, is associated with a jump in the order parameters. This jump is also found in equilibrium treatments using Bose dynamical mean field theory (B-DMFT) \cite{Byczuk17} or quantum Monte Carlo (QMC) \cite{Batrouni17}, and is corroborated by a hysteretic behavior found in the experiments \cite{Landig16, Esslinger18}. This justifies our cluster mean-field treatment also in the presence of dissipation. We determine the regions of coexistence associated with the order-parameter discontinuities for the homogeneous and self-organized atomic phases, namely the MI and DW, SF and DW as well as the SF and SS phases. Hence, within the coexistence region, the corresponding phases are bistable.

To investigate this bistability in more detail, we adiabatically tune the atom-photon coupling $\lambda$ (or, equivalently, the pump rate) across the coexistence region and find hysteretic behavior, a hallmark of the bistability.  
Furthermore, we study the time-dependent out-of-equilibrium dynamics, particularly in the SF-SS bistability region. We prepare the initial state away from any of the bistable saddle points and observe persistent switching behavior between the two coexisting atomic condensate phases. Including cavity-photon loss, we investigate the resulting dissipative dynamics using the Lindblad master-equation approach.
In the presence of cavity-photon loss, the coexistence region is shifted to higher values of the coupling strength $\lambda$, and the switching dynamics are damped so that the system relaxes to one of the two stable attractors characterized by different steady states in the long-time limit. 

The plan of the paper is as follows. We introduce the Dicke-Bose-Hubbard model and describe the cluster mean-field method used to describe the atomic phases in Sec.~\ref{model}. In Sec.~\ref{PD-ground-state}, we compute the ground-state phase diagram, followed by calculating the bistability regions in Sec.~\ref{bistability}. We discuss the non-equilibrium dynamics within the coexistence region, which includes hysteresis dynamics as well as dynamically switching between the coexisting phases in Sec.~\ref{sec-hysteresis} and in Sec.~\ref{oscillatory}, respectively. Finally, in Sec.~\ref{dissipation}, we study the dissipative dynamics and discuss the steady states to which the system relaxes in the long-time limit. We summarize our results and conclude in Sec.~\ref{conclusion}.

\section{The Dicke-Bose-Hubbard model}

\subsection{The model and method}
\label{model}

\begin{figure}[t]
\centering
\includegraphics[width=0.6\columnwidth]{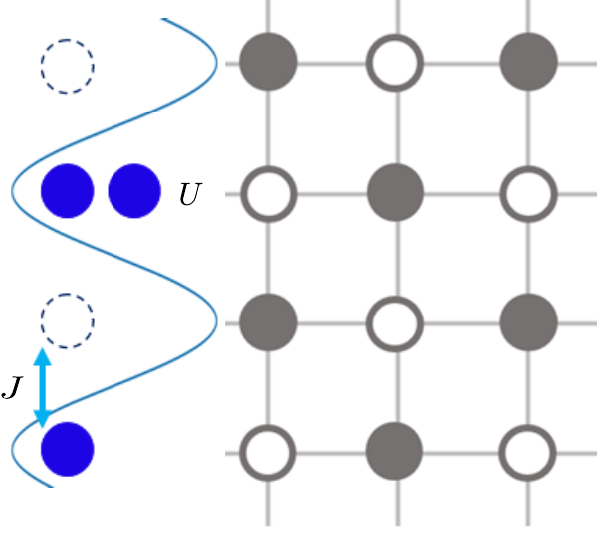}
\caption{{\bf Schematic of the model.} 2D optical lattice with even (open circles) and odd (filled circles) sites, representing $V_{\vec{r}}=\pm 1$, respectively. The dynamical potential generated by the cavity-photon field is shown by the blue curve. The Hubbard parameters $J$ and $U$ in the Hamiltonian \eqref{ham_DBH} are schematically demonstrated.}
\label{fig_Model}
\end{figure}

The dynamics of interacting bosons in a two-dimensional optical lattice coupled to a single mode of an optical cavity can be described by the Dicke-Bose-Hubbard model \cite{Landig16, Kollath20},
\begin{eqnarray}
\hat{H} &=& \hat{H}_{\rm c} + \hat{H}_{\rm b} + \hat{H}_{\rm bc} \label{ham_DBH}\\
\hat{H}_{\rm c} &=& \delta \,\hat{a}^\dagger \hat{a} \nonumber \\
\hat{H}_{\rm b} &=& -J \sum_{\langle \vec{r}, \vec{r}^{\prime}\rangle} \hat{b}_{\vec{r}}^{\dagger} \hat{b}_{\vec{r}^{\prime}}^{\phantom{\dagger}} -\mu \sum_{\vec{r}} \hat{n}_{\vec{r}} +\frac{U}{2} \sum_{\vec{r}} \hat{n}_{\vec{r}}\left(\hat{n}_{\vec{r}}-1\right) \nonumber\\
\hat{H}_{\rm bc} &=& -\frac{\lambda}{\sqrt{L}} (\hat{a}+\hat{a}^\dagger) \sum_{\vec{r}}V_{\vec{r}}\,\hat{n}_{\vec{r}} \nonumber
\end{eqnarray}
where, $\hat{H}_{\rm c}$ describes the cavity mode in the rotating frame with detuning $\delta=\Omega-\Omega_{\text{L}}$ with respect to the laser-pump frequency $\Omega_{\text{L}}$, with $\hat{a}$ ($\hat{a}^\dagger$) the annihilation (creation) operators of the photon field. The bosonic atoms in an optical lattice alone are described by the Bose-Hubbard Hamiltonian $\hat{H}_{\rm b}$, where $\hat{b}_{\vec{r}}^{\phantom{\dagger}}$  ($\hat{b}_{\vec{r}}^\dagger$) and $\hat{n}_{\vec{r}}= \hat{b}_{\vec{r}}^{\dagger} \hat{b}_{\vec{r}}^{\phantom{\dagger}}$ represent the local annihilation (creation) and number operators of bosonic atoms, respectively, at the 2D lattice coordinates $\vec{r} \equiv (i,j)$. $J$ is the hopping amplitude of atoms between the nearest neighbor (NN) sites denoted by $\langle \vec{r}, \vec{r}^{\prime}\rangle$, $U$ is the onsite interaction strength and $\mu$ is the chemical potential. Due to the atom-cavity coupling, the cavity-mode amplitude acts as a potential for the atoms. In the 2D Dicke model, the lattice spacing of the optical lattice is engineered to be half the wavelength of the optical cavity mode in both of the two spatial directions (see Fig.~\ref{fig_Model}). This implies a staggered potential described by $\hat{H}_{\rm bc}$, where $V_{\Vec{r}}=(-1)^{(i+j)}$ and $\lambda$ denotes the potential strength, which is directly controlled by the intensity of the external laser pumping the atomic excitations. Since the photon number in the (extensive) cavity mode scales with the total number of sites $L$ as $a^{\dagger}a\sim L$, the local coupling of the cavity mode to the atoms on site $\vec{r}$ implies the factor $1/\sqrt{L}$ in $\hat{H}_{\rm bc}$ in order for the entire Hamiltonian to be an extensive quantity. 

In the mean-field treatment of the photonic part of the Hamiltonian \eqref{ham_DBH} it is assumed that all cavity photons are in a phase-coherent state, so that the photon field operator $\hat{a}$ can be replaced by its average value, neglecting non-coherent photon fluctuations \cite{Rosch21}. Then the coupling Hamiltonian $\hat{H}_{\rm bc}$ contains only atomic operators and, hence, the total density matrix $\hat{\rho}_{\rm T}$ factorizes into density matrices acting only in the cavity-photon or in the atomic Hilbert space, respectively,  $\hat{\rho}_{\rm T}=\hat{\rho}_{\rm c}\otimes \hat{\rho}_{\rm b}$. Atom-photon correlations thus factorize as 
$\langle \hat{a} \hat{n}_{\vec{r}}\rangle = \langle \hat{a}  \rangle \langle\hat{n}_{\vec{r}}\rangle$, where 
$\langle (\dots)\rangle={\rm Tr} [\hat{\rho}_{\rm c, b}(\dots)]$ denote the thermal average in the cavity and the atomic Hilbert subspace, respectively. In particular, the SR order parameter is defined as the photon amplitude per site, $\alpha={\rm Tr} \left(\hat{\rho}_{\rm c} \hat{a}\right)/\sqrt{L}$ in the thermodynamic limit, $L\to\infty$  and the average occupation imbalance per site as,  $\Delta = \sum_{\vec{r}} V_{\Vec{r}} \langle\hat{n}_{\vec{r}}\rangle/L$ which will appear in the SR phase, see below. 

The mean-field solution of the atom-cavity system at zero temperature is then obtained by minimizing the average cavity energy per site  
\begin{equation}
    E(\alpha,\Delta) = \langle \hat{H}_{\rm c+bc}\rangle/L = \delta |\alpha|^2 - 2\lambda {\rm Re}[\alpha]\Delta,
    \label{cavity-energy}
\end{equation} 
with respect to $\alpha$ for given $\Delta$, where the cavity-photon Hamiltonian is $\hat{H}_{\rm c+bc}=\hat{H}_{\rm c}-\frac{\lambda}{\sqrt{L}} (\hat{a}+\hat{a}^\dagger) L\Delta$, and solving the atomic Hamiltonian 
\begin{eqnarray}
\hat{H}_{\rm b+bc} &=& \hat{H}_{\rm b} - 2\lambda {\rm Re}[\alpha] \sum_{\vec{r}}V_{\vec{r}}\hat{n}_{\vec{r}}
\label{H_MF}
\end{eqnarray}
for given SR field $\alpha$. This process is iterated selfconsistently towards convergence. 

\begin{figure}[t]
\centering
\includegraphics[width=\columnwidth]{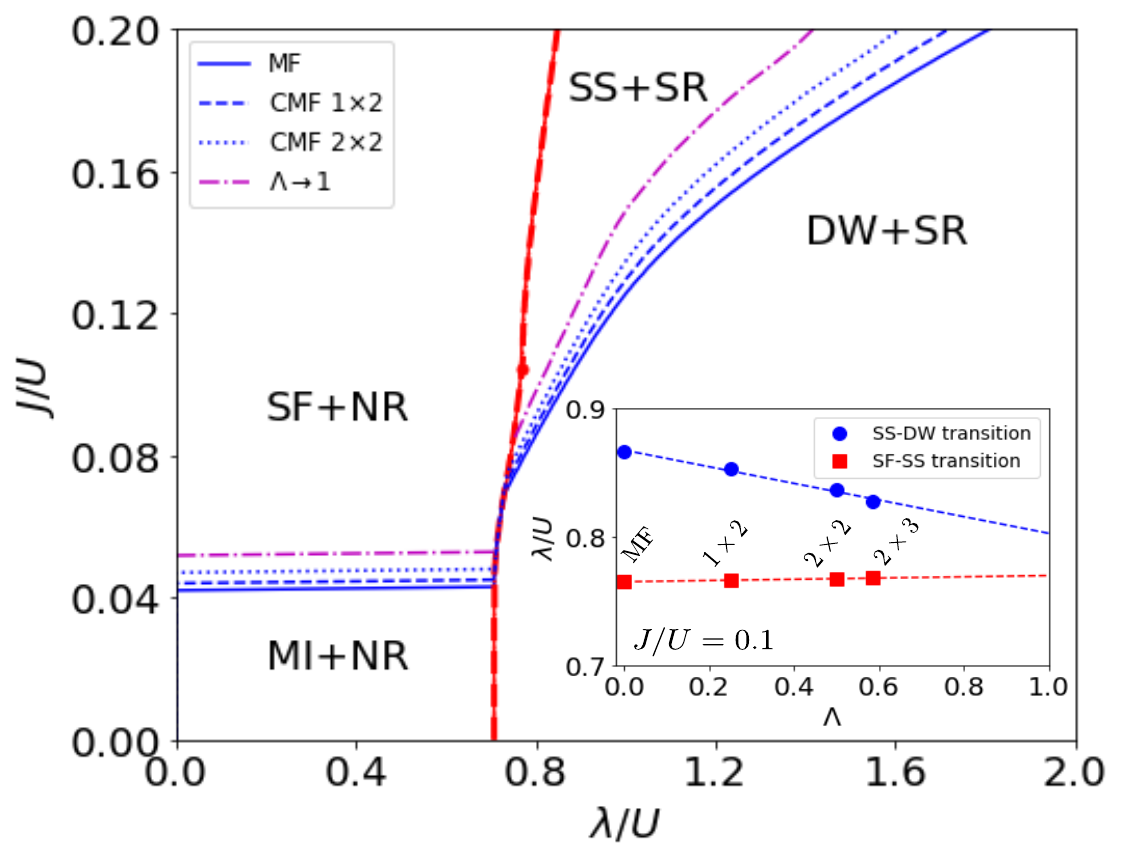}
\caption{{\bf Equilibrium phase-diagram.} Phase diagram for average atomic density per site, $\bar{n} = 1$, for vanishing dissipation $\kappa/U=0$. The red line represents the Dicke transition from the normal-radiant (NR) to superradiant (SR) state of photons. The blue lines represent the atomic condensation transitions from MI and DW to SF and SS phases, respectively. The solid lines are obtained using the single-site mean-field (MF) theory on a bipartite lattice. The phase boundaries obtained using the cluster mean-field theory for various cluster sizes are shown by dashed $(1\times 2)$ and dotted $(2\times 2)$ lines. The extrapolation to infinite cluster size is depicted by the dashed-dotted line.  Examples of this extrapolation, $\Lambda \rightarrow 1$ \cite{Luehmann13,Pohl22}, are shown in the inset for the SF-SS (red squares) and the SS-DW (blue circles) phase boundaries. The Dicke transition line is nearly independent of the cluster size used.}
\label{PD}
\end{figure}

Solving $\hat{H}_{\rm b+bc}$ for the matter fields, we apply the Gutzwiller cluster mean-field theory (CMF) on a two-dimensional, bipartite Bose-Hubbard lattice   \cite{Luehmann13,Pohl22}. Here, the entire lattice is decomposed into smaller $m\times n$-sized clusters. For any cluster $\mathcal{C}_l$, the atomic field operator $\hat{b}_{\vec{r}},~\vec{r} \notin \mathcal{C}_l$, in the neighboring cluster is approximated by its average value $\langle \hat{b}_{\vec{r}}\rangle$, i.e., the local condensate amplitude. The total Hamiltonian \eqref{H_MF} can, thus, be written as a sum of cluster Hamiltonians, $\hat{\mathcal{H}}_{\rm b+ac} = \sum_l \hat{\mathcal{H}}_{\mathcal{C}_l}$, where the $l$th cluster Hamiltonian reads,
\begin{eqnarray}
\hat{\mathcal{H}}_{\mathcal{C}_l} &=& -J\left[\sum_{\langle \vec{r},\vec{r}^{\prime}\in \mathcal{C}_l \rangle} \hat{b}_{\vec{r}}^{\dagger}\hat{b}_{\vec{r}^{\prime}}^{\phantom{\dagger}} + \sum_{\langle \vec{r}\in \mathcal{C}_l,\vec{r}^{\prime}\notin \mathcal{C}_{l}\rangle} \hat{b}_{\vec{r}}^{\dagger}\langle\hat{b}_{\vec{r}^{\prime}}^{\phantom{\dagger}}\rangle + \text{h.c.} \right] \nonumber \\
&+& \sum_{\vec{r} \in \mathcal{C}_l} \left[\frac{U}{2} \hat{n}_{\vec{r}}(\hat{n}_{\vec{r}}-1) - \mu \hat{n}_{\vec{r}} - 2\lambda {\rm Re}[\alpha] V_{\vec{r}}\hat{n}_{\vec{r}} \right]
\label{CMF_ham}
\end{eqnarray}
Accordingly, the total atomic density matrix  factorizes as, $\hat{\rho}_{\rm b} = \prod_l \hat{\rho}_{\mathcal{C}_l}$.
The cluster Hamiltonian $\hat{\mathcal{H}}_{\mathcal{C}_l}$ includes the local condensate amplitudes on the neighboring cluster sites which, due to translation invariance, can be computed as zero-temperature expectation values on the respective sites within the cluster $\mathcal{C}_l$. Hence, the cluster Hamiltonian \eqref{CMF_ham} is diagonalized selfconsistently with the neighboring clusters, and the resulting atomic density imbalance $\Delta$ is inserted into the selfconsistency loop of Eqs.~\eqref{cavity-energy}, \eqref{H_MF}.  

\subsection{Zero-temperature phase diagram}
\label{PD-ground-state}

We first review the zero-temperature phase diagram for this atom-photon coupled system in the $\lambda/U - J/U$ plane in the limit of zero dissipation, analogous to  previous theoretical studies for systems in equilibrium with the cavity-field \cite{Byczuk17, Batrouni17, Donner16, Morigi19, Simon22, Landi18, Yi21}.

\begin{figure}[t]
\centering
\includegraphics[width=\columnwidth]{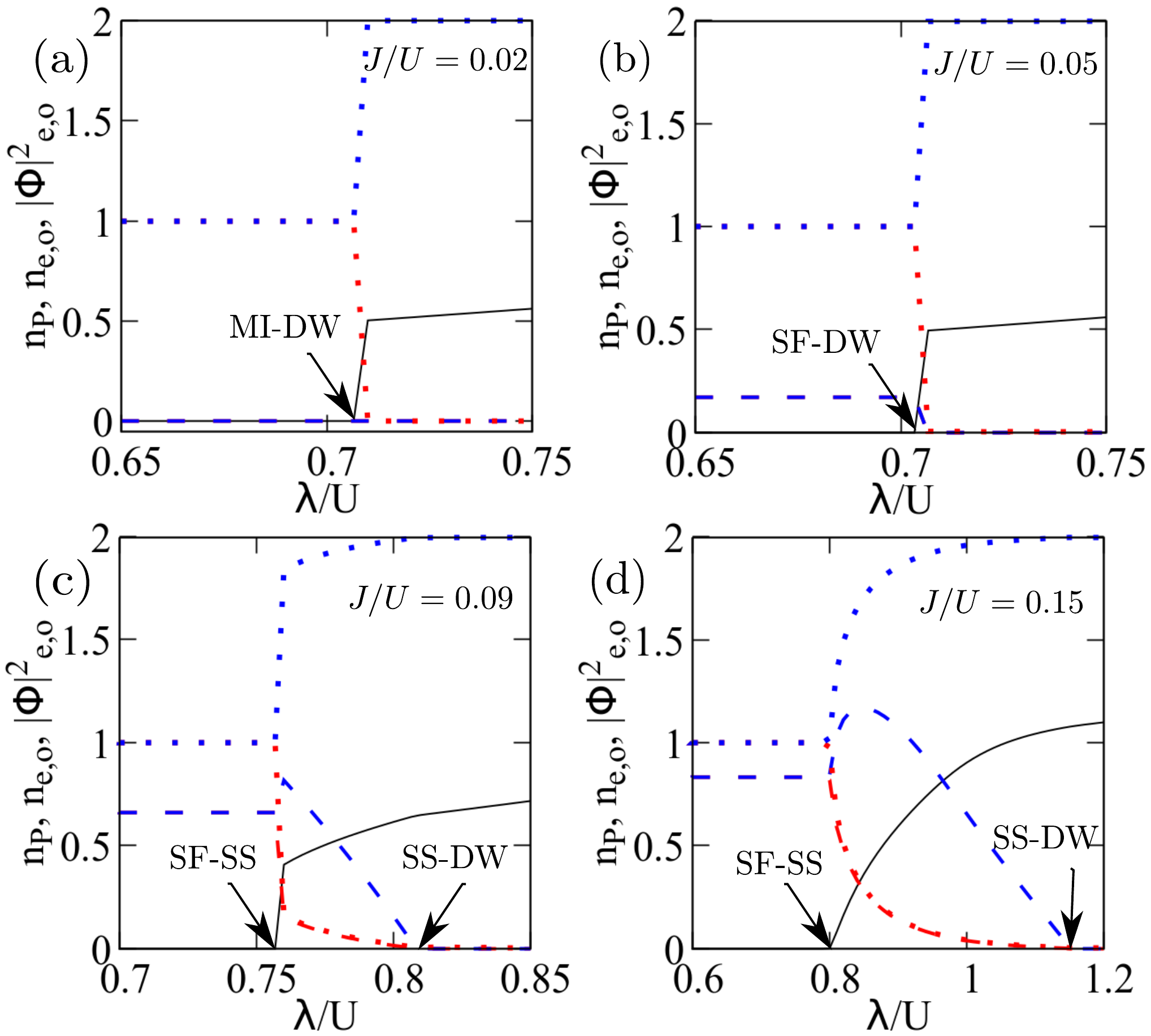}
\caption{{\bf Characterization of phases from orderparameters.} The photon number density $n_p=|\alpha|^2$ (solid black line) and the atomic normal and condensate densities, $n_{\rm e (o)}$ (dotted lines) and $|\Phi|^2_{\rm e (o)}$ (dashed lines), respectively, for the even (odd) sites are plotted as a function of atom-cavity coupling strength $\lambda/U$ for $J/U = 0.02$ (a), $0.05$ (b), $0.09$ (c), and $0.15$ (d). Here and in the subsequent figures, the even (odd) sites are marked by blue (red) colors. Below the critical point $(J_c/U,\lambda_c/U) \approx (0.105,0.77)$, the Dicke transition is accompanied by a jump in all the orderparameters.
}
\label{OP}
\end{figure}

In the normal-radiant (NR) phase of the cavity mode,  defined by the cavity-photon density vanishing in the limit of infinite system size as an order parameter (OP), $n_{\text{P}}=|\alpha|^2 \stackrel{L\to\infty}{\longrightarrow} 0$, the atom system can be in the Mott insulating (MI) or a Bose-Einstein condensed superfluid (SF) phase, depending on the lattice tunneling strength $J$. At the Dicke transition, i.e., when the cavity-mode becomes superradiant (SR) with increasing pump strength $\lambda/U$, $|\alpha|^2>0$, the atomic density becomes spatially modulated due to the staggered potential in $\hat{H}_{\text{bc}}$. 
\begin{figure}[t]
\centering
\includegraphics[width=\columnwidth]{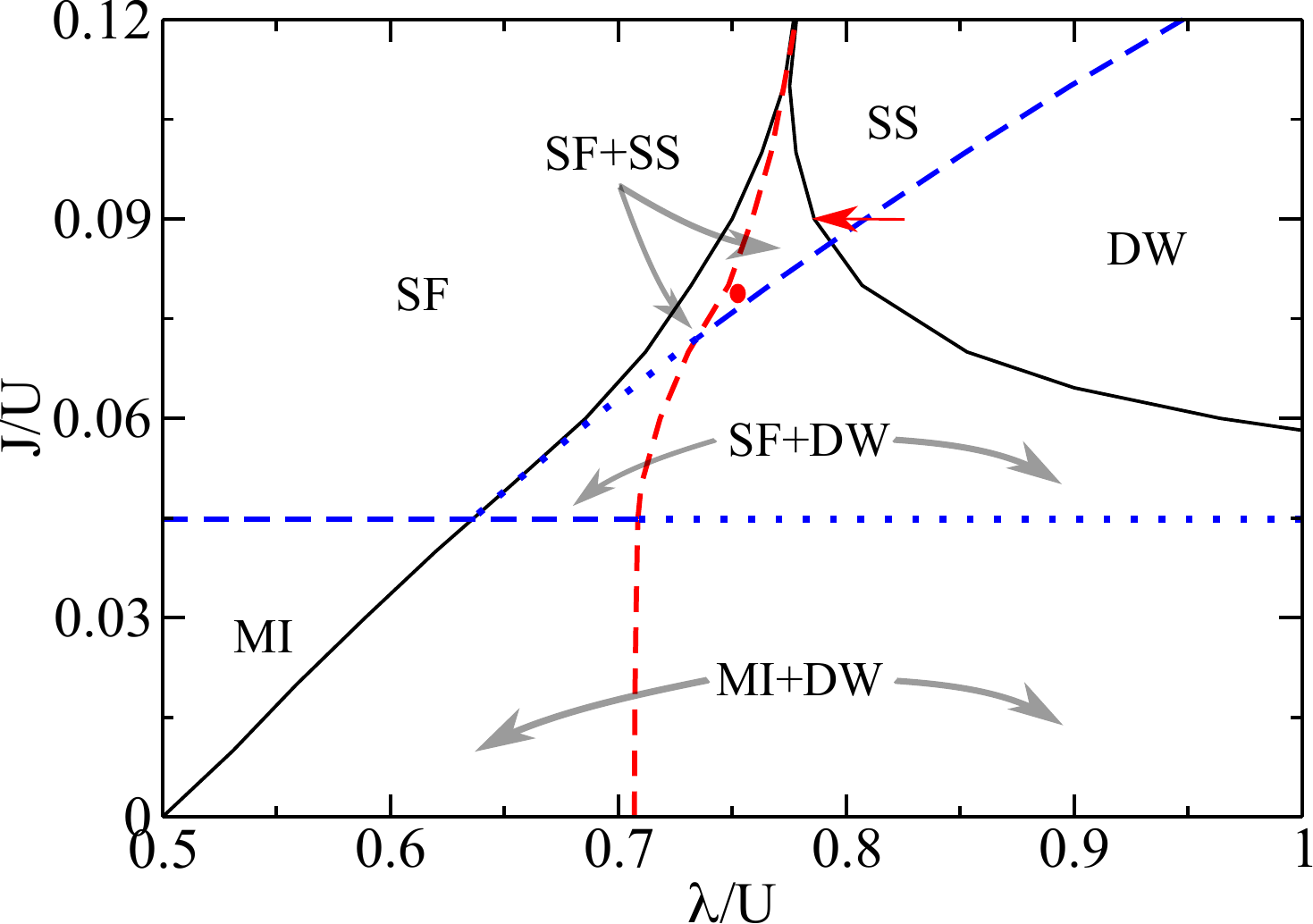}
\caption{{\bf Bistability and coexistence of phases.} Coexistence regions between homogeneous (MI and SF) and self-organized (DW and SS) atomic phases, which appear with normal-radiant and superradiant photon phases, respectively, are depicted in the $\lambda/U-J/U$ plane. The equilibrium transitions in Fig.~\ref{PD} are plotted as dashed lines for guidance. The solid lines mark the boundaries of the coexistence regions. The dotted lines represent the extensions of equilibrium transitions (shown as dashed lines) to the regions where the respective phases are metastable. The blue (red) lines indicate continuous (discontinuous) nature of the transitions at the boundaries. The red filled circle and the red arrowhead represent the region of phase diagram explored during dynamics in Fig.~\ref{hysteresis} and Fig.~\ref{dynamics-bistability}, respectively.
}
\label{PD-bistability}
\end{figure}
Consequently, the phase transitions occur from MI to an insulating density wave (DW), from the SF to a simultaneously Bose-Einstein condensed and spatially modulated state, i.e., a supersolid (SS).    
The combination of these photonic and atomic states yields the four phases shown in the phase diagram in Fig.~\ref{PD}. The atomic subsystem is solved using cluster mean-field theory for different cluster sizes. This yields phase boundaries between the atomic phases (MI-SF, SS-DW) depending on the cluster size as seen in Fig.~\ref{PD}, where the extrapolation to infinite cluster size is shown in the inset (see \cite{Luehmann13,Pohl22} for details) and corresponds to the dashed-dotted lines in Fig.~\ref{PD}.
Note that, since the cluster mean-field theory is applied to the
atomic subsystem, not to the photon subsystem, it does practically not affect the Dicke superradiant transition line, while it does change the transition lines in the atomic subsystem.
We find that the zero-temperature Dicke transition, shown as the red line in Fig.~\ref{PD}, is discontinuous for $J/U$ below the critical point $(J_c/U,\, \lambda_c/U)\approx (0.105,\,0.77)$ marked by the red dot, see Fig.~\ref{OP}. This is consistent with previous studies in equilibrium 
\cite{Byczuk17, Batrouni17, Donner16, Morigi19, Simon22, Landi18,Yi21}, including B-DMFT \cite{Byczuk17} and QMC \cite{Batrouni17}.
Note also that the phase diagram exhibits a tricritical point where the SF+NR, DW+SR, and SS+SR phases meet.       
Representative cuts through the phase diagram for several values of $J/U$, displaying the photonic OP $n_{\text{P}}$ and the atomic OPs, normal density $n_{\text{e,o}}$  and condensate density $|\Phi|^2_{\text{e,o}}$ as functions of $\lambda/U$, are shown in Fig.~\ref{OP}. The subscripts e and o denote the densities on even and odd sites, respectively. The modulated DW and SS phases are characterized by $n_{\text{e}} \neq n_{\text{o}}$ and $|\Phi|^2_{\text{e}}\neq |\Phi|^2_{\text{o}}$, respectively. These results show the OP discontinuities at the Dicke superradiant transition below the critical point $(J_c/U,\, \lambda_c/U)$, i.e., at the MI-DW, SF-DW, and SF-SS transitions, while the atomic condensation transitions, DW-SS and MI-SF, are continuous.        

\begin{figure}[t]
\centering
\includegraphics[width=\columnwidth]{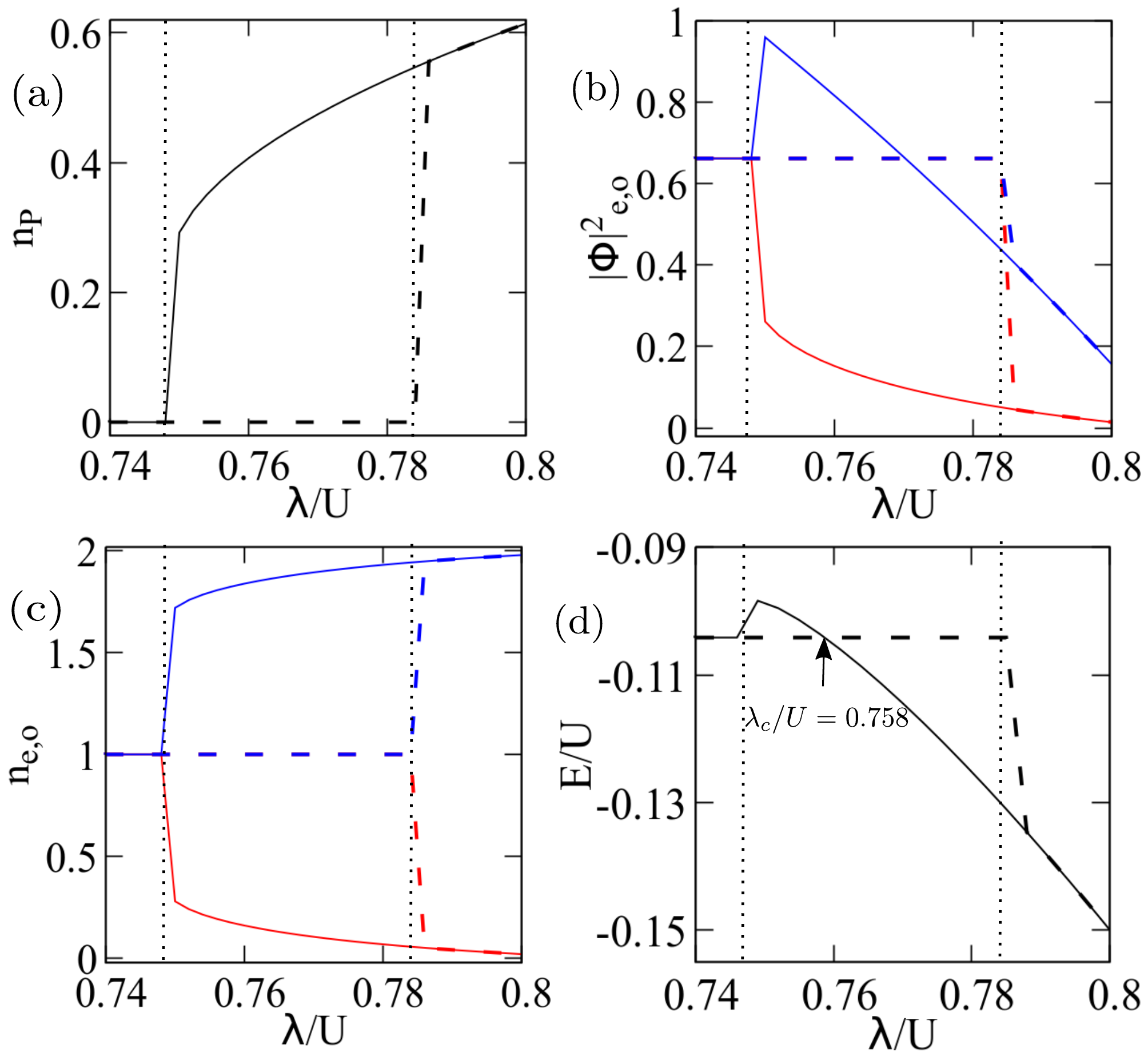}
\caption{{\bf Orderparameters in SF-SS bistability.} The photon number density $n_p$ (a), the atom densities $n_{\rm e,o}$ (b) and atomic-condensate densities $|\Phi|^2_{\rm e,o}$ (c) in the even (blue) and odd (red) sites, and the average energy per site $E/U$ (d) are plotted as a function of the atom-cavity coupling strength $\lambda/U$ for $J/U = 0.09$. The solid and dashed lines are obtained after convergence in the MF self-consistency starting from two different initial states, see the text for details. The vertical dotted lines denote the bistability boundaries as in Fig.~\ref{PD-bistability}. In (d), the two energy lines cross each other at $\lambda_c/U=0.758$ (marked by the arrowhead), indicating the Dicke transition as shown in Fig.~\ref{PD},~\ref{OP}.   
}
\label{bistability-OP}
\end{figure}

\subsection{Bistability and co-existence}
\label{bistability}

At a discontinuous phase transition, one expects a coexistence region of the neighboring phases, corresponding to two minima of the (free) energy as function of the order parameters \cite{Wouter20, Valag21, Valag22}. Our goal is to explore this discontinuous behavior in the mean-field Dicke transition, occurring due to an interplay between the strong atomic interaction and the atom-cavity coupling, from the time-dependent dynamics.
In this section, we investigate the coexisting phases of atoms as well as of photons. In order to converge to a specific one of the coexisting phases, we choose as the initial state of the selfconsistent iterations (c.f., Sec.~\ref{model}) a solution deep inside that phase.

In Fig.~\ref{PD-bistability}, the equilibrium transition lines are shown by the dashed lines as in Fig.~\ref{PD}, for reference. While, the boundaries of the resulting coexistence regions are shown by the solid lines. With increasing hopping, we observe three different coexistence regions, namely the MI+DW, SF+DW and SF+SS, as shown by the arrows. The continuous MI-SF transition, which to the left of the Dicke transition (red dashed line) is the ground-state transition, extends into the superradiant side as a transition between metastable states. The dotted lines in Fig.~\ref{PD-bistability} mark these metastable transitions. The two bistability boundaries meet the Dicke transition line at the critical point $(J_c/U,\, \lambda_c/U)$, above which the Dicke transition becomes continuous [cf. Fig. \ref{OP}~(d)].

As an example, the coexistence between the SF and the SS phases is demonstrated for $J/U=0.09$ by plotting the corresponding OPs in Fig.~\ref{bistability-OP}(a-c). In Fig.~\ref{bistability-OP}~(d), we have also plotted the average energy-density per site of the coupled atom-photon system $E = \langle \hat{\mathcal{H}}_{\mathcal{C}_l} \rangle/N_{\mathcal{C}} + \delta |\alpha|^2$. Outside of the bistability boundaries (thin, vertical lines in Fig.~\ref{bistability-OP}), the OPs and the energy density $E$ converge to the unique solution, however, within the bistability region they lead to different solution with different energies. The two energy curves cross each other at $\lambda_c/U=0.758$, as indicated by the arrowhead in Fig.~\ref{bistability-OP}~(d), representing the ground-state Dicke transition.

\section{Out-of-equilibrium dynamics}
\label{dynamics-neq}

To this end, we study the non-equilibrium dynamics of this atom-photon interacting system. The dissipative dynamics of the system in the presence of cavity-photon loss with rate $\kappa$ can be described within the Markovian approximation by the Lindblad master equation,
\begin{equation}
i\dot{\hat{\rho}}_{\rm T}(t) = \left[\hat{H},\hat{\rho}_{\rm T}(t)\right] + \kappa \left(\hat{a}\hat{\rho}_{\rm T}(t)\hat{a}^\dagger - \frac{1}{2} \left\{\hat{a}^\dagger \hat{a} ,\hat{\rho}_{\rm T}(t) \right\} \right)
\label{eq_diss}
\end{equation} 
where, $\hat{\rho}_{\rm T}(t)$ is the time-evolved total density matrix. Within the mean-field treatment of atom-photon coupling (cf. Sec.~\ref{model}), $\hat{\rho}_{\rm T}(t)$ is factorized into density matrices, $\hat{\rho}_{\rm c}(t)$ and $\hat{\rho}_{\rm b}(t)$, of the cavity-photon and bosonic atoms, i.e. $\hat{\rho}_{\rm T}(t)=\hat{\rho}_{\rm c}(t)\otimes \hat{\rho}_{\rm b}(t)$. Accordingly, time evolution of $\hat{\rho}_{\rm c}(t)$ and $\hat{\rho}_{\rm b}(t)$ are, respectively, governed by,
\begin{eqnarray}
i\dot{\hat{\rho}}_{\rm c}(t) &=& \left[\hat{H}_{\rm c+bc},\hat{\rho}_{\rm c}(t)\right] + \kappa \left(\hat{a}\hat{\rho}_{\rm c}(t)\hat{a}^\dagger - \frac{1}{2}\{\hat{a}^\dagger \hat{a} ,\hat{\rho}_{\rm c}(t)\}\right) \nonumber \\
i\dot{\hat{\rho}}_{\rm b}(t) &=& \left[\hat{H}_{\rm b+bc},\hat{\rho}_{\rm b}(t)\right]
\label{CMF-dyn}
\end{eqnarray}
where, $\hat{H}_{\rm c+bc}$ and $\hat{H}_{\rm b+bc}$ are the cavity-photon and atomic Hamiltonian in Eq.~\ref{cavity-energy} and Eq.~\ref{H_MF}, respectively. The semiclassical equation of motion of the photon-amplitude, namely $\dot{\alpha}(t)=\langle \hat{a}\dot{\hat{\rho}}_{\rm c}(t)\rangle/\sqrt{L}$, leads to
\begin{equation}
\dot{\alpha}(t)=-(i\delta+\kappa/2)\alpha(t) -i\lambda \Delta(t)
\label{cavity}
\end{equation}
where, $\Delta(t)$ is obtained from the atomic dynamics. We perform CMF method in the atomic Hamiltonian (cf. Sec.~\ref{model}) leading to $\hat{\mathcal{H}}_{\rm b+ac} = \sum_l \hat{\mathcal{H}}_{\mathcal{C}_l}$ and $\hat{\rho}_{\rm b} = \prod_l \hat{\rho}_{\mathcal{C}_l}$. Thus, in a cluster, $\hat{\rho}_{\mathcal{C}_l}(t)$ is evolved under $\hat{\mathcal{H}}_{\mathcal{C}_l}(t)$ as,
\begin{equation}
\dot{\hat{\rho}}_{\mathcal{C}_l}(t) = -i[\hat{\mathcal{H}}_{\mathcal{C}_l}(t),\hat{\rho}_{\mathcal{C}_l}(t)]
\label{atom}
\end{equation}
It can be noted that $\hat{\mathcal{H}}_{\mathcal{C}_l}(t)$ depends on the mean-fields, namely time-evolved cavity field $\alpha(t)$ and the condensate amplitude on neighboring cluster $\langle \hat{b}_{\vec{r}}\rangle=\text{Tr}(\hat{\rho}_{\mathcal{C}_l}(t)\hat{b}_{\vec{r}})$ (cf. Eq.~\ref{CMF_ham}). Thus, the coupled Eqs.~\eqref{cavity} and \eqref{atom} are time-evolved to study the dynamics of the system. 

\begin{figure}[t]
\centering
\includegraphics[width=\columnwidth]{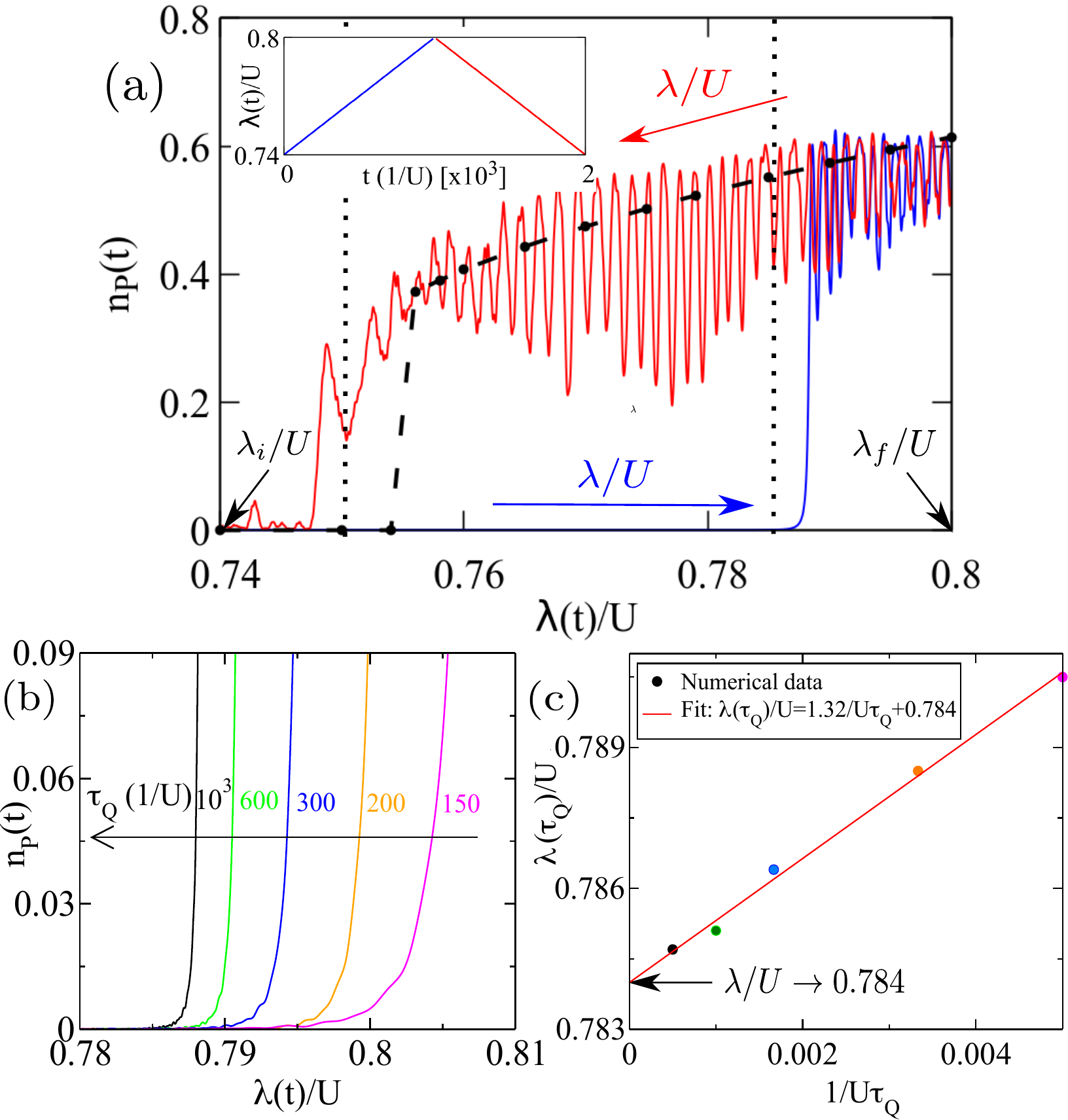}
\caption{{\bf Hysteretic dynamics.} (a) Time evolution of photon number $n_{\rm P}(t)$ is shown as a function of $\lambda(t)/U$, which is linearly ramped up from $\lambda_i/U=0.74$ to $\lambda_f/U=0.8$ (marked by the arrowheads) with a ramp time of $U\tau_{Q}=1000$ and then ramped down with the same rate (see inset). The vertical dotted lines indicate the boundary of the bistability region for $J=0.09$, see Fig.~\ref{PD-bistability}. The photon number obtained from the equilibrium calculation is plotted as the dashed line and has a discontinuous jump at the Dicke transition. In (a) and in Fig.~\ref{hysteresis}~(a,b), we plot the temporal envelope of the dynamical quantities averaged over the time-period $T=2\pi/\delta$ corresponding to fast $\delta$-oscillations. (b) The onset of finite $n_{\rm P}(t)$ near upper bistability boundary is shown for several ramp times $U\tau_Q$ and the corresponding critical couplings $\lambda(\tau_Q)/U$ are plotted in (c) as a function of $1/U\tau_Q$. The linear fitting of the numerical data (see the text for details) shown as solid red line yields the upper bistability boundary in the limit of $1/U\tau_Q\rightarrow 0$ (marked by the arrowhead).    
}
\label{hysteresis}
\end{figure}

We first investigate the time evolution for vanishing dissipation leading to unitary dynamics with $\kappa/U=0$. In particular, we study the hysteresis as well as the switching dynamics between the two co-existing phases when the initial state is chosen away from one of the saddle points, especially in the `SF+SS' bistability region.

\subsection{Hysteretic dynamics}
\label{sec-hysteresis}

We fix the hopping strength at $J/U=0.09$ and prepare the initial state for dynamics as the ground state at $\lambda/U=\lambda_i/U=0.74$. The coupling is then ramped up linearly from $\lambda_i/U=0.74$ to $\lambda_f/U=0.8$ over a time duration $U\tau_Q$, followed by ramped down again to $\lambda_i/U=0.74$ in the same time duration. The ramp time, $U\tau_Q$, is chosen sufficiently large such that the corresponding rate of change in coupling is much smaller than the energy difference of the co-existing phases near the bistability boundaries [cf. Fig.~\ref{bistability-OP}~(d)].
The resulting dynamics of the photon number, $n_{\rm P}(t)$, is shown in Fig.~\ref{hysteresis}~(a). We clearly see a hysteretic behavior, a footprint of bistability. Since the SF-SS coexistence region lies in the interval of $\lambda/U \in [0.75,0.78]$ (dotted lines), the system follows the SF saddle point during ramp up (blue curve) with $n_{\rm P} \sim 0$ until near the upper bistability boundary $\lambda/U \sim 0.78$, where the system moves to oscillate around the unique SS saddle point with finite values of $n_{\rm P}$. The oscillation continues on ramping down (red curve) until it hits the lower SF-SS bistability boundary, where the system becomes SF again. 
For comparison, we have plotted the photon number in the equilibrium as a function of $\lambda/U$ as in Fig.~\ref{OP}~(c), which exhibits a jump at the Dicke-transition, $\lambda_{\rm c}/U \approx 0.757$. The other OPs, such as condensate and atomic densities (not shown here), also exhibit similar hysteretic behaviour as in Fig.~\ref{hysteresis}~(a). 

The dependence of the dynamical transition, particularly the upper bistability boundary, on ramp-time $U\tau_Q$ is demonstrated in Fig.~\ref{hysteresis}~(b,c). The onset of finite $n_{\rm P}$ during ramp-up is shown in Fig.~\ref{hysteresis}~(b) for several values of $U\tau_Q$. With increasing ramp time, the growth of $n_{\rm P}$ occurs at a smaller coupling, which in the adiabatic limit, that is, $1/U\tau_Q\rightarrow 0$, corresponds to the upper bistability boundary as in Fig.~\ref{bistability-OP}. This is clearly shown in Fig.~\ref{hysteresis}~(c), where the coupling strength $\lambda(\tau_Q)/U$ corresponding to the onset of finite $n_{\rm P}$ is plotted with $1/U\tau_Q$. The numerical data are fitted with a linear function, $\lambda(\tau_Q)/U=a/U\tau_Q+b$, with the fitting parameters, $a$ and $b$. The fitted value $b=0.784$ represents the coupling strength, where the transition occurs in the limit $1/U\tau_Q\rightarrow 0$. This agrees with the upper bistability boundary as marked by arrowhead in Fig.~\ref{hysteresis}~(c).

\begin{figure}[t]
\centering
\includegraphics[width=\columnwidth]{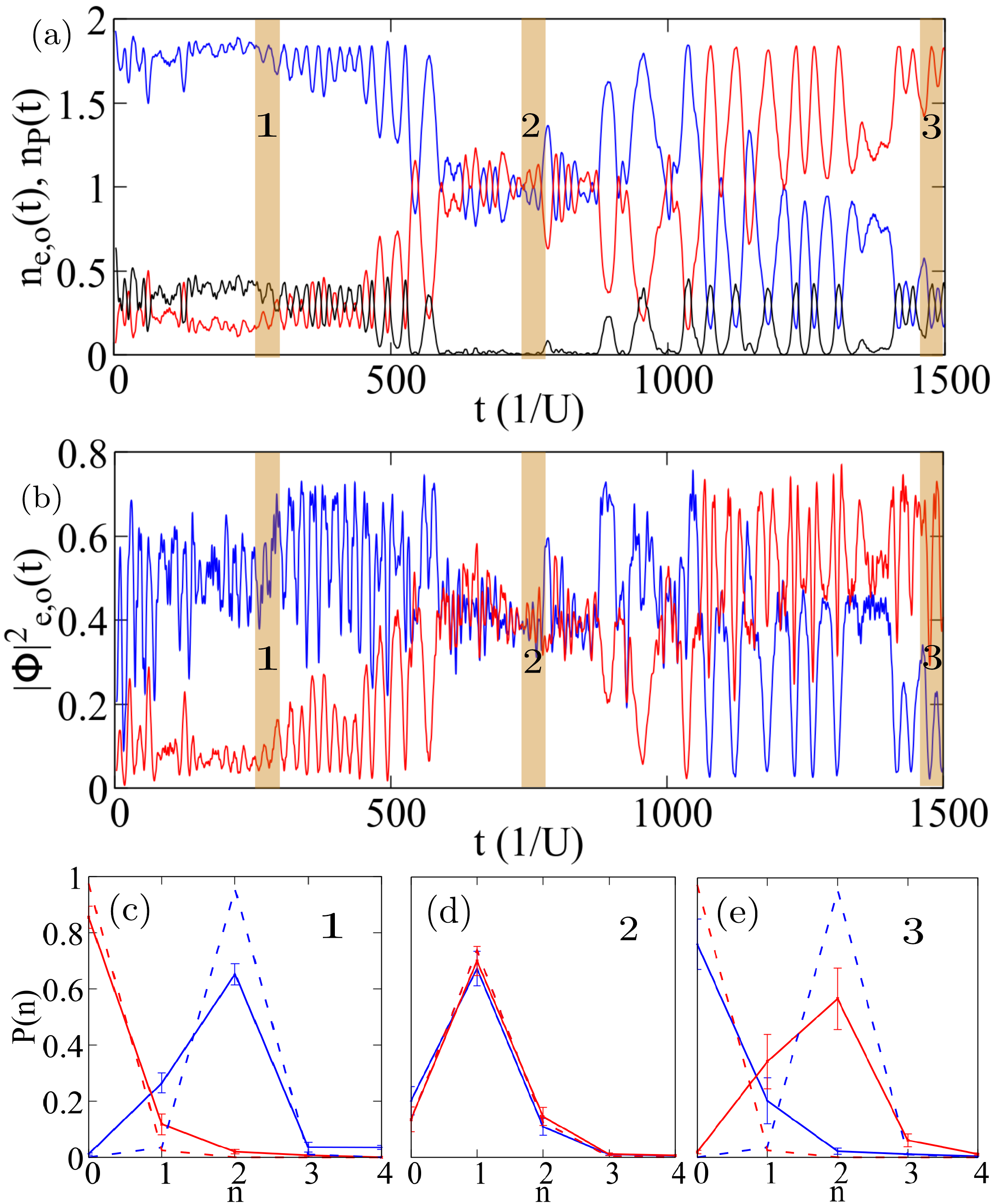}
\caption{{\bf Switching dynamics.} The time dependence of the physical observables (a) photon number $n_{\rm P}(t)$ (black line) and atom densities $n_{\rm e, o}(t)$ as well as (b) the condensate densities $|\Phi_{\rm e, o}(t)|^2$ (blue and red lines) is shown for $\lambda/U=0.76$ and $J/U=0.08$. It is seen that the system stochastically alternates between superradiant, modulated and normal-radiant, homogeneous states. (c-e) Occupation number distributions $P(n)$ of atoms in even (blue) and odd (red) sites at the three different times marked as 1, 2, 3 in (a,b). The distributions are time-averaged over the time window shown as the vertical bars in (a,b). The error bars indicate the variance corresponding to the temporal fluctuations. The dashed lines represent the $P(n)$ distributions obtained from the self-consistent minimization in equilibrium (see Sec.~\ref{bistability} for details).   
}
\label{dynamics-bistability}
\end{figure}

\subsection{Switching dynamics between coexisting phases}
\label{oscillatory}

In this section, we aim to demonstrate the coexistence of phases from an out-of-equilibrium dynamics. We fix the parameters at $J=0.08$ and $\lambda=0.76$, where the system has bistability between SF and SS phases with SS the ground state of the system. Out-of-equilibrium dynamics is performed by preparing the initial state at time $t=0$ in the SS state, however, the corresponding superradiant-photon field is set away from the saddle point $\bar{\alpha}$ by $\delta \alpha$, that is, $\alpha(t=0)=\bar{\alpha}+\delta \alpha$. Thus, the system is driven out-of-equilibrium by providing additional energy larger than the energy difference between the co-existing phases, such that the system can also switch to the other metastable phase.  

The resulting dynamics is elucidated in Fig.~\ref{dynamics-bistability}. In (a,b) we plot time-evolution of the photon-number density $n_{\rm P}(t)$ and the local atomic and condensate densities, $n_{\rm e, o}(t)$ and $|\Phi|^2_{\rm e, o}(t)$, respectively, at even and odd sites. The photon number exhibits oscillations around a finite value for certain time, while, the atomic and condensate densities in the even and odd sites oscillate around different values indicating the SS phase. Whereas, for some time-window, photon number goes down to zero, and correspondingly, the atomic densities in both sites oscillate around the uniform average density, $\bar{n}=1$. Thus, the system dynamically switches between SS and SF phases. Due to the $\mathcal{Z}_2$ symmetry of the Hamiltonian, the photon phase can change dynamically from $0$ to $\pi$. As a result, the atomic OPs switch between the even and odd sites as clear from time domains `1' and `3'. 

In order to investigate the time-evolved atomic phase in detail at different time-windows, we compute the number-state distribution $P(n)$ of atoms locally at each site of the $1\times 2$ cluster. From the time-evolved density matrix of the cluster $\hat{\rho}_{\mathcal{C}_l}(t)$, we construct the reduced density matrix for each of the sites, namely $\hat{\rho}_{\rm e (o)}={\rm Tr}_{\rm o (e)}\hat{\rho}_{\mathcal{C}_l}(t)$. The number-state distribution for any of the sites is then given by, $P(n)=\langle n | \hat{\rho}_{\rm e, o} | n \rangle$. In Fig.~\ref{dynamics-bistability}(c-e) we have plotted $P(n)$ for the even and odd sites in different time-windows marked in Fig.~\ref{dynamics-bistability}~(a). For comparison, we have also plotted $P(n)$ corresponding to the SS and SF phases of the bistability obtained in the self-consistent calculation in equilibrium. In the time-windows marked as `1' and `3', the $P(n)$ distribution is in close resemblance to the SS phase up to the phase of the photon field leading to the switching of $P(n)$ distribution between the even (blue) and odd (red) lattice sites in Fig.~\ref{dynamics-bistability}~(c) and Fig.~\ref{dynamics-bistability}~(e), respectively. On the other hand, in the time window marked as `2', the $P(n)$ distribution resembles the homogeneous SF phase, see Fig.~\ref{dynamics-bistability}~(d). 

\begin{figure}[t]
\centering
\includegraphics[width=\columnwidth]{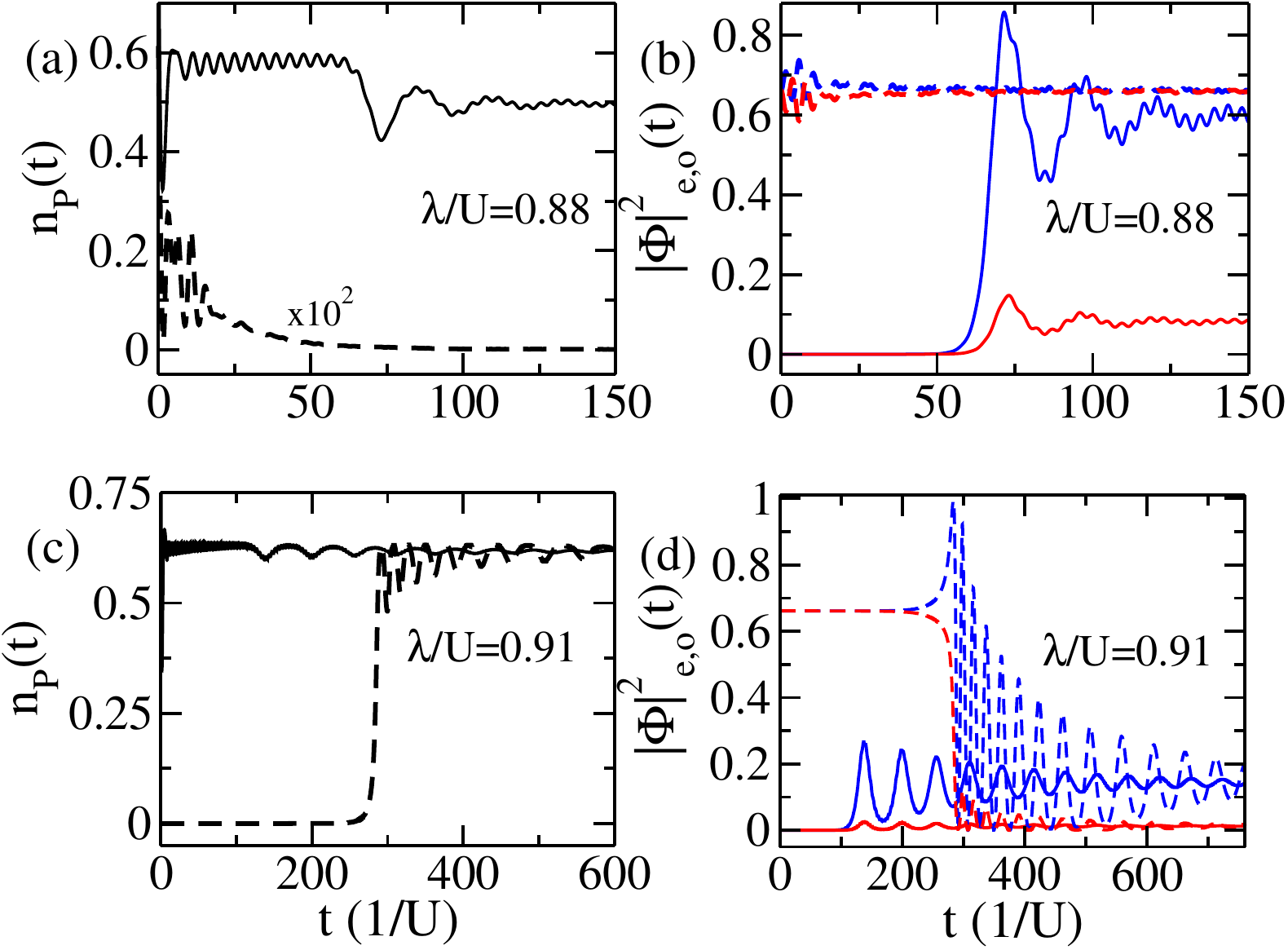}
\caption{{\bf Dissipative dynamics.} The photon number $n_{\rm P}(t)$ (black lines) and condensate-atom densities in the even (blue) and odd (red) sites, $|\Phi|^2_{\rm e, o}(t)$ are plotted as a function of time for atom-cavity coupling strength $\lambda/U=0.88$ (a,b) and $\lambda/U=0.91$ (c,d) with $J/U = 0.09$ and $\kappa/U=1.08$. The time evolution obtained from the initial SF and DW are depicted by dashed and solid lines, respectively. 
}
\label{OP_dynamics}
\end{figure}

\section{Dissipative dynamics}
\label{dissipation}

Finally, we study the dissipative dynamics of the system by time-evolving the coupled equations, Eqs. \eqref{cavity} and \eqref{atom}, for a finite rate of photon loss $\kappa/U$. Apart from characterizing the steady states emerging in the long-time limit, we investigate fate of the observed bistability in the presence of dissipation. For this purpose, we prepare two different initial states, namely, a homogeneous SF and a self-organized DW at $t=0$ in the dynamics.

\begin{figure}[t]
\centering
\includegraphics[width=\columnwidth]{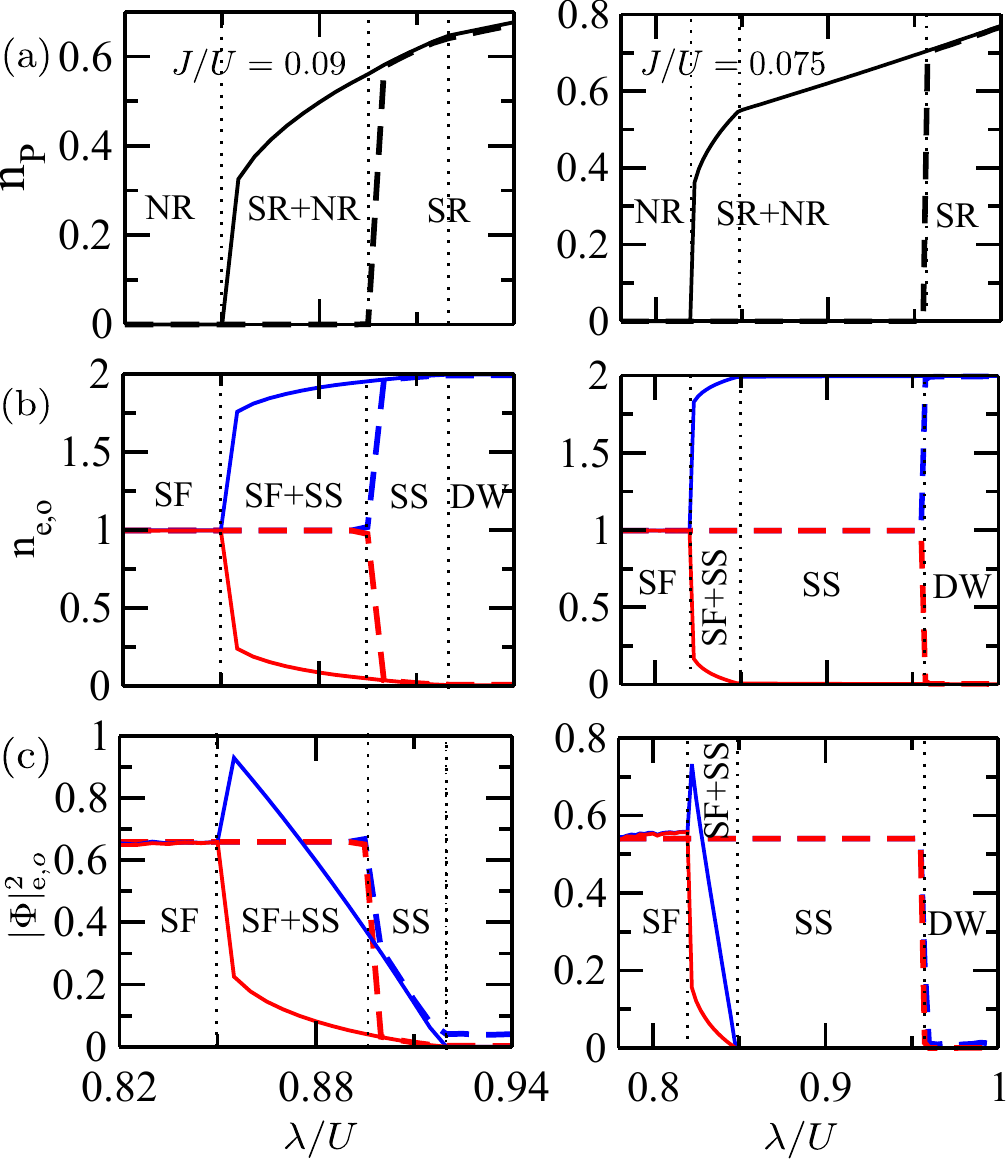}
\caption{{\bf Steady-states in the long-time limit.} The order parameters in the steady states, in particular, (a) the photon number, $n_{\rm P}$, and (b,c) the atom and condensate-atom densities, $n_{\rm e, o}$ and $|\Phi|^2_{\rm e, o}$, respectively, in the even (blue) and odd (red) sites are plotted as a function of $\lambda/U$ for $J/U=0.09$ (left panel) and $J/U=0.075$ (right panel) with $\kappa/U=1.08$ demonstrating the bistability between SF-SS (NR-SR) and SF-DW (NR-SR) phases, respectively. The solid and dashed lines represent the steady states obtained by starting from two different initial choices of states as explained in Fig.~\ref{OP_dynamics}.}
\label{OP_diss}
\end{figure}

A typical time-evolution of the orderparameters, namely the photon-number density $n_{\rm P}$ as well as the condensate-atom densities, $|\Phi|^2_{\rm e, o}$, in the even and odd sites are depicted in Fig.~\ref{OP_dynamics} for two different coupling strengths. Contrary to the time evolution shown in Fig.~\ref{dynamics-bistability}, the system, in the presence of dissipation, relaxes to a steady-state in the long-time limit. In Fig.~\ref{OP_dynamics}, we have compared the time-evolution of $n_{\rm P}$ and $|\Phi|^2_{\rm e, o}$ for the two aforementioned different initial choices. Accordingly, in the region of co-existence, the two initial states relax to the SF and SS fixed points, respectively [cf. Fig.~\ref{OP_dynamics}~(a,b)], indicating the presence of bistability between the SF and SS phases for the chosen coupling strength, $\lambda/U=0.88$. However, for a larger coupling, $\lambda/U=0.91$, the two initial states relax to the unique SS fixed point indicating the absence of bistability, see Fig.~\ref{OP_dynamics}~(c,d).

To determine the coexistence region, the values of the orderparameters in the steady states obtained in the long-time limit are plotted in Fig.~\ref{OP_diss}. In particular, we have demonstrated the SF-SS and SF-DW, or correspondingly NR-SR, phase coexistences in the presence of dissipation. In comparison to Fig.~\ref{PD-bistability} and \ref{bistability-OP}, the coexistence regions are shifted towards a larger coupling strength in the presence of dissipation, however, width of the bistability window remains finite. 

\begin{figure}[t]
\centering
\includegraphics[width=\columnwidth]{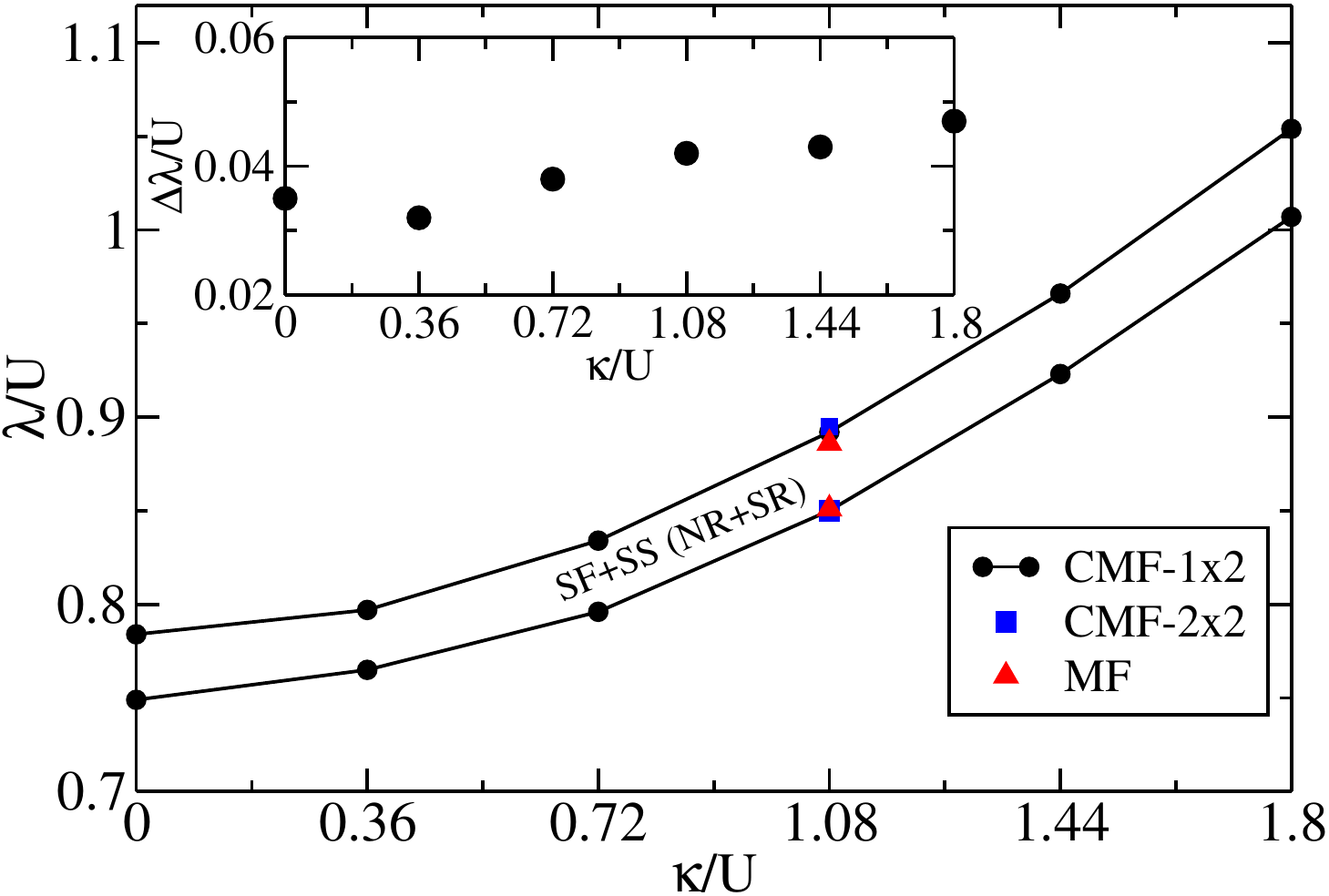}
\caption{{\bf Bistability in the presence of dissipation.} The SF-SS (NR-SR) coexistence boundaries are plotted in the $\lambda/U-\kappa/U$ plane for $J/U=0.09$. Note that the data at $\kappa/U=0$ corresponds to the coexistence boundaries in Fig.~\ref{PD-bistability}. The inset shows the width of the bistability region, $\Delta\lambda/U$, as a function of $\kappa/U$. For comparison, the bistability boundary at $\kappa/U=1.08$ is computed for different cluster sizes, as indicated in the inset.}
\label{PD-dissipation}
\end{figure}

In Fig.~\ref{PD-dissipation}, we have plotted the boundaries of the SF+SS or the NR+SR coexistence region computed at $J/U=0.09$ for various decay rates, $\kappa/U$. In the inset of Fig.~\ref{PD-dissipation} we have shown width of the bistability region in terms of the coupling strength, namely $\Delta \lambda/U$, as a function of $\kappa/U$. 
Clearly, the bistability region occurs at a higher coupling in the presence of dissipation, however, $\Delta \lambda/U$ remains less sensitive to the dissipation rate.  

\section{Conclusion}
\label{conclusion}

In summary, we have investigated the phases and non-equilibrium dynamics of an atomic Bose gas in a two-dimensional optical lattice coupled to a single mode of an optical cavity described by the Dicke-Bose-Hubbard model. The zero-temperature equilibrium phase diagram exhibits four distinct phases, the homogeneous Mott insulator and the Bose-Einstein condensed atomic phases in the photonic normal-radiant state, while in the superradiant state the atomic density is spatially modulated, leading to a density-wave or a supersolid phase, depending on the atomic tunneling strength $J/U$. We find within mean-field theory that the Dicke normal-to-superradiant transition is discontinuous below a critical tunneling strength $J_{\rm c}/U$, consistent with experiments \cite{Landig16,Esslinger18} and with Bose dynamical mean-field theory \cite{Byczuk17} and quantum Monte Carlo \cite{Batrouni17} studies. We mapped out the coexistence regions of normal- and superradiant phases or, equivalently, of homogeneous and modulated atomic phases on both sides of the Dicke transition and found hysteretic behavior upon adiabatically ramping the pump parameter $\lambda/U$ through these coexistence regions. Within the coexistence regions, the time-dependent non-equilibrium dynamics shows stochastic switching behavior between the bistable phases. Upon introducing dissipation described by the Lindblad master equation, this switching dynamics becomes damped and relaxes into the stable state for the given initial parameters of the system. 

We note that non-condensed density fluctuations \cite{Rosch21} in both, the cavity-mode and the atomic systems, beyond mean-field theory may lift the discontinuity at the Dicke transition. It is, however, to be expected that this critical fluctuation region is more narrow than the coexistence region mapped out in our work. This means that away from the transition line two minima of the free energy with different minimal values should exist on both sides of the transition, so that the temporal switching behavior will persist and should be experimentally observable outside of the fluctuation-dominated, critical region.

\section*{Acknowledgements}

The authors thank H. Ritsch, M. Thorwart, M. Fleischhauer and M. Kajan for useful discussions. This work was funded by the Deutsche Forschungsgemeinschaft (DFG) under Germany’s Excellence Strategy-Cluster of Excellence Matter and Light for Quantum Computing, ML4Q (No. 390534769) and through the DFG Collaborative Research Center CRC 185 OSCAR (No. 277625399). S.R. acknowledges a scholarship of the Alexander von Humboldt (AvH) Foundation, Germany.

\end{document}